# The Impact of the Global Crisis on SME Internal vs. External Financing in China

ShiXue He[1,*] and Marcel Ausloos[1,2,**]

[1]School of Business, University of Leicester
Leicester, LE1 7RH, United Kingdom

[2]GRAPES,
Group of Researchers for Applications of Physics
in Economy and Sociology,
rue de la Belle Jardinière, B-4031 Liège,
Federation Wallonie-Bruxelles, Belgium

- email: heshixuewh@gmail.com
  **corresponding author: ma683@le.ac.uk, marcel.ausloos@ulg.ac.be

**Abstract**

Changes in the capital structure before and after the global financial crisis for SMEs are studied, emphasizing their financing problems, distinguishing between internal financing and external financing determinants. The empirical research bears upon 158 small and medium-sized firms listed on Shenzhen and Shanghai Stock Exchanges in China over the period of 2004-2014. A regression analysis, along the lines of the Trade-Off Theory, shows that the leverage decreases with profitability, non-debt tax shields and the liquidity, and increases with firm size and tangibility. A positive relationship is found between firm growth and debt ratio, though not highly significantly. It is shown that the SMEs with high growth rates are those which will more easily obtain external financing after a financial crisis. It is recognized that the China government should reconsider SMEs taxation laws.

 E41  Demand for Money
 C25  Discrete Regression and Qualitative Choice Models

 G01  Financial Crises
 C55  Large Data Sets: Modeling and Analysis

I.     INTRODUCTION

The financing sources of small and medium size enterprises (SMEs) can be divided into internal financing and external financing (Ross et al., 2013). The internal financing is obtained from firm owners, retained earnings or depreciation (Ross et al., 2013).

By itself, internal financing cannot satisfy an SME development. Thus, SMEs are looking for other means of financing: among these, bank loans are among the main sources. However, banks set several restrictions when lending to SMEs; for example, banks increase the costs for loans as well as the collateral and shorten the repayment period. Often, SMEs cannot provide enough collateral assets or reliable financial statements to offset the information asymmetry and adverse selection risks for money lenders (Paulet et al., 2014). Understandably, banks prefer to deal with large, old, known companies with high information transparency (Nguyen et al., 2015). Thus, compared to internal financing, external financing is expensive and hard to obtain for small businesses (Jiang et al., 2014).

One has already been much concerned about the impact of the financial crisis on SMEs, - mainly due to the consequently numerous bankruptcies. One can point to the overall financial environment of a country or to the whole world. A practical cause has been found in the increased cost of production together with a decreased demand of whatever product. It is easily pointed out that the decrease in profit implies that any internal financing possible options will decrease.

Moreover, the global financial crisis which broke out in 2008 is considered by the IMF (*http://www.theguardian.com/business/2008/apr/09/useconomy.subprimecrisis*) to have been the most dangerous crisis since the Great Depression. During this recession period, the financing problems for SMEs became even worse than usual. For example, the banks especially large banks set several restrictions when lending to SMEs: banks increase the costs for loans as well as the collateral and shorten the repayment period in order to reduce risks during the financial crisis. Consider the PR China case: according to the statistics of CBRC (China Banking Regulatory Commission), the total lending by state-controlled banks in 2008 are 2.2 trillion. However, the small business loans only account for 15 percent (i.e. 300 billion). More than 20 percent of the registered SMEs went bankrupt and another 20 percent are still facing severe shortages of capital, e.g. as noticed in the first quarter of 2009 (Cunningham, 2011). Compared to the state-owned businesses, the SMEs received much less protection and support from the China government during the crisis

Thus, it seems useful to pin point the determinants of SMEs capital structure changes, in particular *ca.* financial crisis time, in order to expect solutions or at least give some advice in order to reduce unfortunate issues. The present article is organized along such concerns about financing difficulties of SMEs, taking into account information asymmetry, relationships between banks and enterprises, and internal limitations within the SMEs' financial system. Based on the pertinent literature, after identifying a few gaps in previous studies, 6 factors are chosen as being the determinants of the capital structure: beside the dependent variable, i.e. total debt to total asset ratio, one has the (i) returns on assets, - its "profitability", (ii) its non-debt tax shields, (iii) its liquidity, measured by the quick ratio, (iv) the size (assets) of the firm, (v) its tangibility, and (vi) a firm growth characteristics, i.e. the operating profit margin. Noticing that profitability and liquidity of SMEs, two internal financing means, are considered to be more important for getting bank loans, - external financing nowadays than before the crisis, the analysis results indicate that the internal financing difficulties for SMEs should be more seriously tackled by political and economic authorities. This should be emphasized at once, because the capital structure decisions of SMEs differ according to the types of firms. Service SMEs' capital structure decisions are closer to the assumptions of Pecking Order Theory and rather removed from those of Trade-Off Theory compared with the case of other types of manufacturing SMEs (Serrasqueiro et al., 2011). Our analysis focuses on the latter type, suggesting to investigate a Trade-Off Theory model.

II.    LITERATURE REVIEW

In this section, we overview the pertinent literature regarding the objectives so outlined. More can be found in e.g. Abdulsaleh and Worthington (2013).

Although this research is based on China data, the international facet should not be neglected. In brief, one may recall that the theories that explain the "to be perfect capital structure of a firm" reach no consistent conclusion (Seifert and Gonenc, 2010), even after Modigliani and Miller (1958) theorem. The latter authors stated that in the absence of taxes, agency costs or other market imperfections, the market value of a firm is not affected by its leverage (Ross et al., 2013). This theory is based on strict conditions such as the absence of taxes, bankruptcy costs and asymmetric information. Later, Modigliani and Miller (1963) added the corporate taxes into the theory and recognized the tax benefits (tax-shields) from interest payments. Thereafter, the trade-off theory (TOT), which includes a trade-off between the tax benefits from debt and financial distress costs, subsequently implies that there is an optimal debt to equity ratio for every firm which helps to balance the debt benefits and the increase in financial risks. This debt to asset ratio naturally obviously becomes the dependent variable to study.

In contrast, Myers (1984) introduced the Pecking Order (POT) which implies that the financial managers prefer to finance new investments through internal financing (retained earnings). Among others, Fama and French (2002) and Shyam-Sunder and Myers (1999) found that the POT can explain the financing choices made by firms. Fama and French (2005) also claimed that the SMEs exposed to the influence of the information asymmetry, are relying heavily on equity financing instead of debt financing. Indeed, small and medium enterprises do not obey the rules of the pecking order due to the information asymmetry (Frank and Goyal, 2003). Chen (2004), cited in Seifert and Gonenc (2010, p.4) reached the same conclusion for the Chinese market, proposing that Chinese firms obey a "new pecking order hypothesis": retained-earnings, equity and long-term debt (Seifert and Gonenc, 2010).

More arguments on using Pecking Order Versus Trade-off theory framework can be found in Sogorb-Mira and López-Gracia (2003): "An Empirical Approach to the Small and Medium Enterprise Capital Structure".

More generally speaking, one of the major reasons for SMEs' financing difficulties is thought to be the information asymmetry. Indeed, the lack of equal information sharing leads to imbalances in the economy, thereby causing moral hazards and adverse selection problems. The imbalance lies both in the providers of funds and the receivers stands. Unlike large listed companies which can access funds from the capital markets, small and medium enterprises' external financing sources are primarily found in banks. However, the weak information share between SMEs and banks limits the SMEs' availability of banks (Irwin and Scott, 2010). Consequently, SMEs are more vulnerable to capital flows than large firms, - especially during a financial crisis (Dong and Men, 2014). On the other hand, the limited financing channels put SMEs in a weak negotiating position with financial institutions. Not only SMEs do not get any preferential term, as compared with large firms, but also SMEs are constrained by several mandatory provisions, such as the offering of collateral and shortening the loan duration (Ang, 1991).

The lack of face to face communication between SMEs and banks also leads to financing difficulties in another aspect. Banks prefer to choose large firms which have audited financial statements and "good governance" rather than small firms, - this in order to reduce expected credit risks.

This leads us to outline the specific explanatory factors of the model, beside the dependent variable.

2.1 Factors which can affect the internal financing of SMEs.

(i) Profitability
Researchers have different opinions on the relationship between the profitability and financial leverage. Titman and Wessels (1988) argued that the profitability is negatively related to leverage in the US market. More recent studies also support this negative relationship analysing various data (Booth et al. (2001) for developing countries, and Wald (1999) for developed countries). It is argued that the large amount of free cash flow weakens the enterprises' control of management (unnecessary spending). Thus, the shareholders would prefer to choose outside creditors to supervise the management when using external financing (Mallin, 2013). In this case, profitable firms tend to have higher leverage. Yet, bank loans are relatively hard to obtain by SMEs, surely in China. Therefore, the enterprises would choose internal financing first. As the profitability increases, the reliance on external financing would decrease gradually. The assumption in this essay leans toward a negative relationship between debt ratio and profitability.

(ii) Non-debt tax shields
According to TOT, the financial leverage has a positive relationship with debt tax shields and a negative relationship with bankruptcy cost. DeAngelo and Masulis (1980, cited in Cheng and Green, 2008) found that not only the debt financing could provide tax shields, but also other expenses, e.g. depreciation and investment tax credits, have tax benefits as well. (These non-debt tax shields are substitutes for the accounting debt tax shield.) However, both Titman and Wessels (1988) and Song (2005) argued that there is no statistically significant relationship between non-tax shields and debt. (There is also an opposite effect on short-term debt and long-term debt.) Quite contrarily, Shahjahanpour et al. (2010) concluded that there is a negative relationship between the non-tax shields and leverage. The argument stems in the consideration that the depreciation level of the SMEs can affect their internal financing ability. A deduction of this depreciation should be an important source of internal funds. In other words, enterprises with higher non-debt tax shields usually prefer to have less debt, - and vice versa.

Thus, considering the above, the assumption is this essay is that the non-tax shields variable is likely inversely associated with the SMEs' debt ratio.

(iii) Liquidity (Quick Ratio)
The effect of the asset liquidity on capital structure has no consistent conclusion: it has both positive and negative influence according to Mouamer (2011). On one hand, enterprises which have higher liquidity may have relatively greater debt ratio,- in order to meet their short-term obligations. On the other hand, enterprises with high liquidity may use these assets to finance their future investment opportunities. Thus, the high liquidity enterprises could borrow less money from the financial institutions. Therefore, one can conclude that there is a negative influence on the enterprises' debt ratio.

In order to calculate the liquidity of the firms, we use the Quick Ratio to measure a company's short-term liquidity. It measures the ability of a company to use its most liquid assets (i.e. current asset minus inventories) to extinguish its current liabilities. As firms' quick ratio increases, the fund utilization rate might increase and the reliance on the debt financing would face a corresponding decrease (De Jong et al., 2008).

Thus, from the above discussion, we hypothesize that liquidity is negatively related to debt ratio.

2.2 Factors which can affect the external financing of SMEs

(iv) Firm Size
Much of the current literature on the financing ability of a firm implies that it is affected by the firm size. It is found that large enterprises usually have relatively higher liabilities (Booth et al., 2001). Abor and Biekpe (2009) also provided evidence that in contrast to small firms, large firms prefer to use debt. The large firms with lower expected bankruptcy costs have relatively more easy access to loans and equity. However, Fama and Jensen (1983) suggest that there is a negative relationship between firm size and debt. The cause for this correlation sign resides in the large firms' tendency to disclose "more information" than small firms. In so doing, the largest firms would be regulated more heavily than small firms, whence limiting the cost of information asymmetry for the former (Abor and Biekpe, 2009).

Based on these assertions, for SMEs, we can assume that the firm size can increase the financing ability of the enterprises and should be positively related to the debt.

(v) Tangibility

As argued by Chen et al. (2013), tangibility (a fixed assets over total assets ratio) is also an essential determinant of capital structure. Related research has shown that due to the information asymmetry, the firm managers can access more secured information on a company than other (external) creditors. Moreover, if the firms use debt financing, agency costs are required. However, the collateral assets would help reducing these costs, whence somewhat the information asymmetry problems. Moreover, the greater the collateral value, the lower the risk for the creditors (Amidu, 2007). Thus, the collateral value can increase the external financing ability of a firm, to some extent.

Thus, under high information asymmetry, financial institutions would fund those enterprises which have higher tangibility. It is therefore hypothesized that there is a positive relationship between tangibility and leverage.

(vi) Firm Growth
In the case of asymmetric information, high growth and competitive enterprises would tend to present a greater external financing ability than otherwise. The more so for SMEs. Nevertheless, under the current economic situation in China (i.e., within the information asymmetry frame), the actual management performance as well as actual economic conditions are hard to be measured by financial institutions. Thus, the firm growth should be considered as an important determinant to be studied.

Heshmati (2001) already found that the fast-growing SMEs tend to have higher leverage, especially in concentrated ownership firms. However, Myers (1977) held the view that high growth firms might give up some investment opportunities, according to its presently positive net value, for various strategic manager incompetence. However, in so doing, this kind of firms' capital structure would have a low proportion of debt, - not withstanding a possible conflict between bond holders and shareholders.

Thus, the relationship between firm growth and debt ratio seems to be an interesting question (Abor and Biekpe, 2009), - a theoretical and empirical gap in the present framework. It is here presently hypothesized that Firm Growth is positively related to the debt ratio.

## III. DATA AND METHODOLOGY

3.1 Sample and data collection
The essay is intended to analyze the small and medium listed enterprises' (SME) financing issue before and after the financial crisis in China. At least 60 percent of China's GDP is created by the SMEs. Meanwhile more than half of the SMEs are in the manufacturing industry sector (Tambunan, 2009). The financing issue represents much of the main SMEs economics problems. One obviously needs some reliable and as much as possible "complete" data. Therefore, due to the difficulty of accessing China SMEs financial data, the data of listed companies is the only way to obtain some coherently meaningful and reliable data.

Thereafter, the data includes 158 manufacturing-listed SME's quarterly reports over 10 years (between January 2004 and December 2014) downloaded from the Guotai database. The subprime crisis period is considered to span from December 2007 to June 2009 as given by the NBER in *http://www.nber.org/cycles/cyclesmain.html*. Their basic statistical characteristics before and after the crisis are given in the Appendix; see Tables A1.1, A2.1, and A2.2. Such data markedly points to differences in statistical characteristics of these SMEs between both time intervals, i.e., before or after the crisis, following a mere visual inspection of the Tables. Further comments are found in the Appendix.

Determinants statistical correlations, based on the Pearson correlation coefficients, are also given in Table A3.1, in the Appendix.

3.2 Dependent Variable
The dependent variable used to determine the financial leverage of a company, in this essay, is the Total Debt to Total Assets ratio, called TDTAR, here below: it indicates how many assets are financed by the debt. It can be used to determine the financial risk of the firms (Sogorb-Mira, 2005). In brief, if the ratio is higher than 1, the company is considered to have problems to pay back the debts and vice versa.

3.3 Explanatory variables
The six explanatory variables used to distinguish between ways of firm financing and its capital structure, before and after the financial crisis have been discussed through the literature review here above: Profitability, Non-debt tax shields and Liquidity, on one hand, Size, Tangibility, and Firm Growth, on the other hand. The codes used to read the model and subsequent tests are given in Table 1.

3.4 Model
The data used in this essay is of the panel data type: it contains both cross-sectional data and time series data. In general, there are two investigation methods for panel data: the random effects model and the fixed effects model (Koop, 2008). The fixed effects model uses dummy variables to model the individual effect and the random effects model do not use dummy variables but assumes that the individual effect is a random variable (i.e. $\varepsilon_{it} = v_i + u_{it}$) (Koop, 2008). This essay employs the so called random effects regression model (Koop, 2008), also in line with Abor and Biekpe (2009) or Hall et al. (2004), since there is no dummy variable; the individual effect is through the random variable $\varepsilon_{it}$. The model is written as

$$TDTAR_{it} = \alpha + \beta_1 ROA_{it} + \beta_2 NDTS_{it} + \beta_3 QR_{it} + \beta_4 SIZE_{it} + \beta_5 TANG_{it} + \beta_6 GROW_{it} + \varepsilon_{it} \qquad (1)$$

where $TDTAR_{it}$ is the firm's debts to assets ratio, i.e., the dependent variable for the $i$ firm at time $t$; $v_i$ and $u_{it}$ stand as a stochastic variable and some error term, respectively. The $\alpha$ and $\beta_i$ coefficients have to be determined.

Table 1. Codes of variables for model and tests.

| Classification | Variable name | Code | Formula |
|---|---|---|---|
| "Internal Financing Variables" | | | |
| Profitability | Return on Assets | ROA | Net Income /Total Assets |
| Non-debt tax shields | Depreciation to Fixed Assets Ratio | NDTS | Depreciation /Fixed Assets |
| Liquidity | Quick Ratio | QR | (Current Assets-Inventories) /Current Liabilities |
| "External Financing Variables" | | | |
| Firm Size | Natural logarithm of total assets | SIZE | ln (total asset) |
| Tangibility | Fixed Assets over Total Assets | TANG | Fixed Assets /Total Assets |
| Firm Growth | Operating Profit Margin | GROW | Operating Income /Net Sales |

4. Empirical Regression Results and Analysis

Table 2. The regression analysis results for the explanatory variables (ROA, NDTS, QR, SIZE, TANG, GROW) and the dependent variable (debt to asset ratio, TDTAR), using 158 SMEs in China, before or after the crisis. Asterisks indicate significance: *** and ** at the 1% and 5% level, respectively.

| TDTAR | Before Crisis | | After Crisis | |
|---|---|---|---|---|
| Variable | Coefficient | Signif. | Coefficient | Signif. |
| ROA | -1.4509 | *** | -2.6061 | *** |
| NDTS | -379.6017 | *** | -205.8583 | ** |
| QR | -0.8930 | ** | -10.6608 | *** |
| SIZE | 119.2079 | *** | 76.3540 | ** |
| TANG | 89.4439 | *** | 24.6262 | ** |
| GROW | -0.0980 | ** | 0.0317 | ** |

The $\beta_i$ coefficients resulting from the regression analysis are given in Table 2 for the 6 explanatory variables (ROA, NDTS, QR, SIZE, TANG, GROW) with respect to the dependent variable (debt ratio: TDTAR) using 158 SMEs in China, distinguishing the before or after the crisis cases.

The regression coefficients are all statistically significant at 5 percent, for both before or after the financial crisis period cases. Moreover, since the coefficients of the variables are finite this means that each variable has an effect on the debt ratio. The 6 variables explain about half of the variation in the dependent variable (i.e. TDTAR, debts to assets ratio) before and after the crisis (0.5074 and 0.4592, respectively), as calculated through the regression coefficient $R^2$.

Furthermore, the F-statistics for both cases are found to be much smaller than 0.05. Thus, one can conclude that there is a significantly positive relationship between TDTAR and the explanatory variables. Notice that the statistical characteristics after the financial crisis (see Appendix) also indicate that the SMEs' capital structure has been influenced by the crisis.

5. Discussion

It seems numerically indubitable, from the above data, that the variables in this regression model do have explanatory powers, allowing us some further theoretical analysis.

First, consider the theoretical factors which practically influence an SME internal financing. (i) The ROA, representing the profitability of the firm, is found to be highly statistically significant. This is consistent with the pecking order theory (Ross et al., 2013): firms with comparatively high profitability would decrease their reliance on external financing and rather use internal financing instead.

(ii) Concerning NDTS, DeAngelo and Masulis (1980) mentioned that the tax deduction advantages of non-debt shields can effectively decrease the firm's debt ratio. However, the non-debt shields (NDTS) coefficients are very negative, in the present cases; in fact, there is a weaker significant impact on the TDTAR after the financial crisis (the p-value is 0.2525). One possible reason for this somewhat surprising fact is conjectured to stem from the government lack of help to the SMEs. In other words, the imperfections of the taxation system in the Chinese market results in a low fixed assets depreciation rate, which thereby causes the concerned firms not to use the depreciation to gain funds. Thus, one deduces that the government policy current taxation system aims to foster the internal financing of the SMEs.

(iii) There is a negative and statistically significant relationship between the liquidity (QR) and the debt ratio, both before and after the crisis, as could be hypothesized. The QR represents the ability for firms to resist risk, measuring a company's ability to meet its short-term obligations. It has decreased, on average, after the crisis (see Tables in Appendix). Nevertheless, the high (in absolute value) quick ratio coefficient value indicates that SMEs could get easily some access to bank loans. However, the (negative values of the) test results contrast with this expectation. Such negative regression coefficient results imply that the liquidity of SMEs is an important factor for risk determination by loan providers, - before and after the crisis. One can propose two possible reasons to explain this finding. First, the quick ratio itself does not provide financial institutions enough confidence on the SMEs. Secondly, the negative correlation implies that the profitability has quite affected their internal financing. In fact, SMEs would give up on expensive external financing if they have sufficient internal funds. To some extent, this result also proves one of our concerns, i.e. the impairing role of information asymmetry between financial institutions and SMEs.

Next, consider the factors which can influence the external financing. There is a positive and statistically significant relationship between the SIZE and the debts to assets ratio TDTAR. Thus, these firm sizes play an important role in determining the capital structure of the firm (Sogorb-Mira, 2005). The bigger the firm size, the more easy one can get bank loans. With increasing, expanding, SMEs sales, the profitability, the quality of the products, and, the more so, the firm credit will increase. Nevertheless, extra funding requirement will increase accordingly. A possible intrinsic mechanism comes in mind: the large firms could use diversified investments to dilute risk. Hence, large firms will have lower bankruptcy cost, as discussed by Titman and Wessels (1988) indeed. On the other hand, large size firms are likely to reveal more information to the public, within some psychological or marketing scheme; in so doing, such an information transparency makes them to appear more reliable than the small firms. Thus, one easily understands that the financial institutions, such as banks, prefer to lend money to large firms rather than to SMEs, - thereby explaining the positive correlations reported in Table A2.

Notice that the impact of the SMEs' tangibility on the leverage after the financial crisis is lower (yet, less significant) than before the crisis. This is interpreted as mainly due to the financial institutions greater awareness of the risks after the crisis. This is a somewhat interesting point to debate in further work.

Finally, the regression results confirm the positive relationship between the firm growth and debt ratio in the SME Chinese market. However, it is less statistically significant. Nevertheless, a theoretical interpretation goes as follows: high growth SMEs have comparatively a more strong desire to expand after than before the crisis. After the crisis, SMEs with high growth rates are likely to obtain more external financing than at times before the financial crisis. The fact that such SMEs resisted better to the crisis is a likely convincing argument for lenders indeed.

In summary, profitability, non-debt tax shields, as well as the liquidity of the SMEs show a negative and significant relationship with the debt ratio. This implies that the profitable SMEs with more liquidity assets prefer to decrease their financial leverage. Secondly, the positive relationships between the financial leverage and both firm size and tangibility suggest that the big SMEs with more fixed assets will prefer more external debt financing. Third, after the financial crisis, the quick ratio and the ROA are more significant than before the financial crisis; this implies that the liquidity and profitability of the SMEs are important determinants for the loan providers, - especially after the crisis. Fourth, the impact of the SMEs' tangibility on the leverage after the financial crisis is less significant than the influence before the crisis. Last but not least, the growth rates and non-debt tax shields of SMEs are not the main factors which influence the firm leverage, especially after the crisis. This is because the debt providers become more cautious due to the high risk of high growth SMEs. The banks tend to be credit grudging due to the information asymmetry (Wehinger, 2014). The SMEs cannot use the depreciation to obtain the internal financing due to the low fixed asset depreciation rate.

5. Conclusion

This research aimed at testing changes of the determinants of small and medium-sized manufacture enterprises capital structure, before and after the financial crisis; 158 SME in China were investigated during the period 2004 to 2014.

We summarize a few undertaken objectives, point their theoretical and practical connections, more specifically focusing on China, but not only, and provide thought for further debates,

(i) To identify how to help the SMEs to achieve rapid development. The argument stems in the common belief that any development of SMEs relates directly to the development of the national economy and creates a large amount of job opportunities. (Moreover, the conference on "Financing SMEs in Europe", in 2008 pointed out that the economic recovery is largely relying on the development of small business; also recall that the small and medium-sized enterprises account for almost 99 percent of the registered enterprises in China; see also successive "Annual Report of SME Finance in China": http://www.smefinanceforum.org/post/china-sme-finance-report-2013).

(ii) To identify a few difficulties which restrain the financing of SMEs and to present corresponding solutions as well as recommendations. There are several reasons which lead to the financing difficulty of the SMEs, - the situation worsening during the financial crisis. Without inserting quantitative means, we nevertheless consider that one of the most serious problems is information asymmetry between the SMEs and capital provider. In other words, the borrower could take advantages of the lender's lack of important information, subsequently resulting in some potential risk. Beside such a frustrating relationship between banks and firms, the limitations within the SMEs' financial system themselves also lead to financing difficulties. The loan decision by financial institutions will be much more dependent on the credit rating. Moreover, the financial institutions are unwilling to lend money to SMEs due to the imperfection of corporate governance structure standardized system and low credit grade. The high bankruptcy rate and high default rate of SMEs during the recession period has made the financing even harder.

On the positive side, the long-term interaction and peer monitoring which are already pointed out by Banerjee et al. (1994), together with the development of small financial institutions could reduce the information asymmetry to some extent. On one hand, it is (of course!) important for SMEs to build a long term relationship with the banks. Thus, the banks could have access to valuable information from (usually opaque) small businesses. On the other hand, small financial institutions do not have as much options as large banks. Therefore, it seems necessary to develop the small financial institutions sector, which might prefer to invest into small business in contrast to large banks. It is indeed often claimed that SMEs' financing channels are insufficient under the current economic environment. Many other alternatives, such as mezzanine financing and financial leasing assets, do recently provide more choice for SMEs indeed.

(iii) To determine the impact of the financial crisis on the SMEs of the Chinese market and the changes of the SMEs' capital structure after the crisis as well as the causes of these changes. More than 20 percent of the SMEs went bankrupt during the financial crisis, and the rest face severe shortages of capital. Moreover, the financing situation for SMEs are even serious due to tighter credit conditions: SMEs are required to provide more collateral assets and to shorten the repayment period. Thus, several determinants of the capital structure have been greatly influenced by the global financial crisis. "Interestingly", in contrast, the recent severe economic crisis was not found to have had an impact on capital structure determinants for Greece SMEs (Balios et al., 2016).

Even though, one might complain that 6 variables are not enough, one could add the number of employees, before and after the crisis, the age of the company, the debt ratio could be divided into long-term debt ratio and short-term debt, the structure of the managerial board, - taking into account gender (Watson, 2006), single or multiple owners (Newman et al., 2013), etc., we consider that the regression results in the present case study prove much for our focus. With respect to the relationship between capital structure determinants and debt, the empirical evidence allows us to draw important conclusions regarding the applicability of the assumptions of Trade-Off Theory to the capital structure decisions of China SMEs at crisis time. In such a framework, our more interesting conclusions are:

    1. The QR and ROA which represent the liquidity and profitability respectively have a significant negative relationship with the debt ratio both before and after the crisis. In other words, the high liquidity which implies a comparatively good risk tolerance ability is important for capital providers under growth and recession cycle period.

    2. The regression analysis results indicate that non-debt tax shield (NDTS) and tangibility (TANG) are not the main determinants which influence the SMEs' capital structure, - in this China case study. We stress that, apparently, the SMEs in China did not take full advantages of the non–debt tax shields, maybe due to the (imperfect or too complex) taxation system. In our it , might be also that they did not get much support from the government, at the recession time.

    3. The firm size (SIZE) and growth (GROW) are factors that influence the external financing of SMEs; both have a s relationship with the debt ratio, - in the Chinese Market before the crisis. However, the P-value of GROW in the regression model has increased to 0.5505. One possible reason is that the high growth SMEs is always accompanied by high risks, due to the information asymmetry itself increasing the financing difficulty. Financial institutions such as banks would be more cautious when they lend money to high growth SMEs during the recession period. The strong positive correlation between the firm size and debt ratio, both before and after the financial crisis, imply that the loan providers believe that large firms have a stronger ability to repay the loans than the small firms. This also an important reason of SMEs financing difficulty: the latter could be reduced if applications of the trade-off theory is optimized internally by SMEs managers, leading to subsequent openness by external lenders. Again, we point toward the need for a reduction of the information imbalance.

In summary, the results of this paper indicate influences of the financial crisis on the SMEs financing difficulty. If a political economy suggestion can be made here, let it be hoped that the Chinese government establishes laws to protect the SMEs as well as to provide more financing channels. It matters (Cotei et al., 2011).

APPENDIX

A.1 Descriptive Statistics

Table A1.1. The skewness and kurtosis of the seven variables value distributions for the examined 158 Chinese SMEs over Jan. 2004 – Dec. 2014

|  | TDTAR | ROA | NDTS | QR | SIZE | TANG | GROW |
|---|---|---|---|---|---|---|---|
| Kurtosis | -0.5917 | 4.1821 | -0.0802 | 46.1743 | -0.0163 | -0.4935 | 48.7021 |
| Skewness | -0.0656 | 1.2536 | 0.6386 | 5.6583 | 0.1484 | 0.2195 | 5.8405 |

Notice that the skewness of the TDTAR distribution is negative; the skewness is positive for the 6 determinants and even greater than 1 for ROA, QR and GROW. Moreover, the excess kurtosis of QR and GROW is quite leptokurtic.

Table A1.2 Descriptive statistics of dependent variable (TDTAR) and
6 explanatory variables of 158 SMEs in China before the financial crisis

| | Before the crisis | | | | | | |
|---|---|---|---|---|---|---|---|
|  | TDTAR | ROA | NDTS | QR | SIZE | TANG | GROW |
| Min | 6.9917 | 0.3969 | 0.00 | 0.1043 | -0.0068 | 0.0069 | -3.9216 |
| Max | 63.9846 | 12.317 | 0.0598 | 13.517 | 0.1561 | 0.5459 | 143.664 |
| Mean | 28.0018 | 4.9934 | 0.0155 | 1.0766 | 0.0427 | 0.1719 | 16.9959 |
| SD | 12.1433 | 2.0788 | 0.0089 | 1.4488 | 0.0281 | 0.0958 | 21.5761 |
| SD/Mean | 0.4337 | 0.4163 | 0.5742 | 1.3457 | 0.6581 | 0.5573 | 1.2695 |

Table A1.3. Descriptive statistics of dependent variable (TDTAR) and
6 explanatory variables of 158 SMEs in China after the financial crisis.

| | After the crisis | | | | | | |
|---|---|---|---|---|---|---|---|
|  | TDTAR | ROA | NDTS | QR | SIZE | TANG | GROW |
| Min | 3.9043 | -3.238 | 0.0043 | 0.0337 | -0.0471 | 0.0469 | -41.1265 |
| Max | 81.5558 | 19.813 | 0.0510 | 6.7322 | 0.1451 | 0.5508 | 307.914 |
| Mean | 42.9165 | 4.3804 | 0.0220 | 0.4430 | 0.0391 | 0.2773 | 13.3546 |
| SD | 17.0988 | 3.2405 | 0.0107 | 0.6938 | 0.0289 | 0.1115 | 31.8104 |
| SD/Mean | 0.3984 | 0.7398 | 0.4864 | 1.5661 | 0.7391 | 0.4021 | 2.3820 |

Comparing the values in Table A1.3 to those in Table A1.2 allows to show that the financial crisis did affect the capital structure of China SMEs.

First, especially for the firms' liquidity: the mean QR indicates that firms had liquid assets available to cover current liabilities before the crisis, but could not do so after the crisis. The QR is one of the most affected terms during the crisis.

The mean TANG (0.1719 to 0.2773) and that of NDTS (0.0155 to 0.0220) have increased after the crisis; this indicates that, after the financial crisis, SMEs have more collateral assets when borrowing money from banks.

In order to have a better understanding of the differences before and after the crisis, one way is to use a ratio, the standard deviation divided by mean, called the coefficient of variation (CV=SD/Mean) (Atkinson and Nevill, 1998). The CV of QR and GROW are the highest both before and after the crisis, which implies that the liquidity and growth of the SMEs are always greatly fluctuating. It is important to notice that the CV of ROA, 0.4163 before the crisis, almost doubled after the crisis (0.7398). The means of the ROA are not very different: 4.9934 before the crisis and 4.3804 after the crisis. However, the SD of the ROA goes up from 2.0788 before the crisis to 3.2405 after the crisis, which implies that the crisis has influenced the profitability spread of the SMEs.

A.2 Correlation analysis

Table A2.1. Pearson correlation coefficient between each variable characterizing the 158 SMEs in the Chinese market.

|      | ROA | NDTS    | QR      | SIZE    | TANG    | GROW    |
|------|-----|---------|---------|---------|---------|---------|
| ROA  | 1   | -0.1857 | 0.5280  | 0.3284  | -0.2540 | 0.7118  |
| NDTS |     | 1       | -0.1632 | -0.3199 | 0.8290  | -0.1851 |
| QR   |     |         | 1       | 0.1121  | -0.2463 | 0.8659  |
|      |     |         |         |         |         |         |
| SIZE |     |         |         | 1       | -0.2421 | 0.2438  |
| TANG |     |         |         |         | 1       | -0.2086 |
| GROW |     |         |         |         |         | 1       |

In order to identify the possibility of multicollinearity among each variable, a correlation matrix of the variables is presented in Table A2.1. The multicollinearity occurs when there are high correlations among explanatory variables (i.e. very close to +1 or -1) which can lead to unreliable and biased results of regression. Except for three correlation statistics visually close to +1, there is no point of debating about a multicollinearity problem (Koop, 2008).